\documentclass[a4paper,12pt]{article}
\usepackage{jheppub} 
\usepackage{lineno}
\newcommand{\llangle}{{\langle\!\langle}}
\newcommand{\rrangle}{{\rangle\!\rangle}}
\newcommand{\csch}{{\mathrm{csch}}}

\arxivnumber{} 

\title{\boldmath R\'{e}nyi Entropy with Surface Defects in Six Dimensions}

\author[a]{Ma-Ke Yuan}
\author[a,b]{and Yang Zhou}

\affiliation[a]{Department of Physics and Center for Field Theory and Particle Physics, \\Fudan University, Shanghai 200433, China}
\affiliation[b]{Peng Huanwu Center for Fundamental Theory, Hefei, Anhui 230026, China}

\emailAdd{mkyuan19@fudan.edu.cn}
\emailAdd{yang\_zhou@fudan.edu.cn}

\abstract{We compute the surface defect contribution to R\'{e}nyi entropy and supersymmetric R\'{e}nyi entropy in six dimensions. We first compute the surface defect contribution to R\'{e}nyi entropy for free fields, which verifies a previous formula about entanglement entropy with surface defect. Using conformal map to $S^1_\beta\times H^{d-1}$ we develop a heat kernel approach to compute the defect contribution to R\'{e}nyi entropy, which is applicable for $p$-dimensional defect in general $d$-dimensional free fields. Using the same geometry $S^1_\beta\times H^5$ with an additional background field, one can construct the supersymmetric refinement of the ordinary R\'{e}nyi entropy for six-dimensional $(2,0)$ theories. We find that the surface defect contribution to supersymmetric R\'{e}nyi entropy has a simple scaling as polynomial of R\'{e}nyi index in the large $N$ limit. We also discuss how to connect the free field results and large $N$ results.}

\begin{document}
\maketitle
\flushbottom
\section{Introduction}
\label{sec:intro}
Defects are important objects in quantum field theory. The familiar examples are Wilson lines and Wilson surfaces. Usually the defects represent some external probes coupled to the theory. For instance a charged heavy probe particle in gauge theory is described by a Wilson line, which plays an important role in understanding the non-perturbative property of the theory. On the other hand, entanglement entropy and R\'{e}nyi entropy have been intensively studied in recent years. They play crucial roles in connecting information theory, field theory and gravity. For instance, they are related to conformal anomalies in conformal field theories and can be computed holographically. For many theories, the R\'{e}nyi entropy, as a one-parameter generalization of entanglement entropy, provides us more information about the theory.

Counting the defect contribution to R\'{e}nyi entropy is therefore of great importance in understanding both the information theoretical structure and the non-perturbative nature of quantum field theory. 

The defect contribution to entanglement entropy has been studied previously in~\cite{Jensen:2013lxa,LM1,Nishioka1,Estes:2018tnu,Gentle:2015jma,Gentle:2015ruo}. Recently the surface defects rekindled much attention~\cite{Henningson:1999xi, Gustavsson:2003hn, Gustavsson:2004gj, Berenstein:1998ij, Graham:1999pm, Chen:2007ir, Chen:2007zzr, Chen:2008ds, Drukker1, Drukker:2020atp, Drukker:2020bes, Drukker:2021vyx, Drukker:2022beq, Drukker:2022kuz, Drukker:2020swu, Gutperle:2019dqf, Gutperle:2020rty, Trepanier:2023tvb, Raviv-Moshe:2023yvq, Giombi:2023dqs, Cuomo:2023qvp, Shachar:2022fqk,Chen:2023lzq,Herzog:2022jqv,Rodriguez-Gomez:2022gbz,Lauria:2020emq,Zheng:2022zkm,Kim:2020npz,Gaiotto:2012xa,Wang:2020xkc}. Although the formula about the surface defect contribution to entanglement entropy has already appeared, for instance in~\cite{Nishioka1,Jensen1}, the way to compute the contribution to R\'{e}nyi entropy is still lacking.

We compute the additional contribution to R\'{e}nyi entropy due to the defects. In particular we focus on conformal surface defects in six-dimensional conformal field theories (CFT). In CFTs, the computation can be mapped to the calculation of thermal entropy with inverse temperature $\beta$ on hyperbolic space. For the purpose of computing entanglement entropy we only need to know the free energy as well as its first derivative with respect to $\beta$ at $\beta=2\pi$. The latter can be expressed as an insertion of stress tensor. However, this is not enough for the purpose of computing the R\'{e}nyi entropy. Instead we have to compute the free energy as a function of $\beta$. Since we only concern the additional contribution due to the defect, we measure the change of the free energy (as a function of $\beta$) due to the defect. The analysis was initiated for line defects in four dimensions by Lewkowycz and Maldacena in~\cite{LM1}. In this note we generalize the analysis to surface defects in six dimensions. The method is applicable for general $p$-dimensional defects in $d$-dimensional free fields. In the entanglement entropy limit, we check against the known results and find precise agreement.

Another motivation to study surface defects in six dimensions is to understand $(2,0)$ theories. This is challenging because the proper formulation of the interacting theories is still lacking. Combining the ideas of supersymmetry and R\'{e}nyi entropy, one can define the supersymmetric refinement of the ordinary R\'{e}nyi entropy for these theories~\cite{Nian:2015xky,Zhou:2015kaj}. The way is to consider $(2,0)$ theories on $S^1_\beta\times H^5$ with an additional R-symmetry background field. Interestingly, the supersymmetric R\'{e}nyi entropy enjoys universal relations with conformal anomalies (as well as 't Hooft anomalies) and can be computed holographically from 2-charge hyperbolic black hole in the large $N$ limit~\cite{Zhou:2015kaj}. Therefore the supersymmetric R\'{e}nyi entropy provides a new observable with an extra parameter for $(2,0)$ theories. Given this development, we study further the most natural operator, the surface operator in $(2,0)$ theories on $S^1_\beta\times H^5$. In particular we compute the surface operator contribution to the supersymmetric R\'{e}nyi entropy and find that it has a simple scaling as polynomial of R\'{e}nyi index in the large $N$ limit.

We also discuss the possible way to connect the free field results and the large $N$ results.

\section{Defects in CFTs}
Let us start by introducing conformal defects in CFTs. We work in Euclidean signature. A $p$-dimensional conformal defect breaks the ambient CFT conformal symmetry $SO(1, d + 1)$ to $SO(1, p + 1) \times SO(d - p)$. For a CFT in flat space, conformal symmetry forces $\langle T_{\mu\nu}(x) \rangle = 0$. However, in the presence of defect, the one-point function of ambient stress-energy tensor does not necessarily vanish. To illustrate, consider a $p$-dimensional planar defect in $\mathbb{R}^d$. The metric is then divided into parallel and transverse directions: 
\begin{equation}
\text{d}s^2 = \text{d} \hat x^a \text{d} \hat x^a + \text{d}x^i \text{d}x^i
\end{equation}
with $a=0, \dots, p-1$ and $i=p, \dots, d-1$. The stress-energy tensor follows from varying the defect partition function and can be split into the ambient part $T^{\mu\nu}$ and the defect localized part $\hat{T}^{ab}$. See~\cite{defect2016,Nishioka1} for instance. The ambient stress-energy tensor is a symmetric traceless tensor of dimension $d$ and spin $2$, hence the (partial) conservation plus residual conformal symmetry fix its form completely~\cite{defect2016}\footnote{The notation $\llangle O \rrangle$ refers to the correlation function measured in the presence of defect
\begin{equation*}
\llangle O \rrangle \equiv \langle O D \rangle / \langle D \rangle\ ,
\end{equation*}
where $D$ denotes the defect. }
\begin{equation}\label{Tbulk}
\begin{split}
\llangle T^{ab} \rrangle &= -\frac{d - p - 1}{d} \frac{h}{|x^i|^d} \delta^{ab}\ , \quad \llangle T^{ai} \rrangle = 0\ ,\\
\llangle T^{ij} \rrangle &= \frac{h}{|x^i|^d} \left(\frac{p + 1}{d}\delta^{ij} - \frac{x^i x^j}{|x^i|^2}\right)\ ,
\end{split}
\end{equation}
where $h$ characterizes the property of the defect.\footnote{In our convention, $h$ in \eqref{Tbulk} corresponds to $-a_T$ in \cite{defect2016}. }
\subsection{Defect anomalies}
Even-dimensional defects have Weyl anomalies, which means that the trace of the defect localized stress-energy tensor does not vanish. For surface defects~\cite{Henningson:1999xi,Schwimmer:2008yh,Graham:1999pm}\footnote{Our normalization differs from~\cite{Jensen1} by a factor of 2.}
\begin{equation}
    \llangle \hat{T}^a_a \rrangle = - \frac{1}{12\pi} \left[ b R^\Sigma + d_1 \Tilde{\Pi}_{ab}^{\mu} \Tilde{\Pi}^{ab}_{\mu} - d_2 W_{ab}^{ab} \right]\ , 
\end{equation}
where $R^\Sigma$ is the intrinsic Ricci scalar of the defect submanifold $\Sigma$, $\Tilde{\Pi}_{ab}^{\mu}$ is the traceless part of the second fundamental form, and $W_{abcd}$ is the pullback of the bulk Weyl tensor to $\Sigma$. $b$, $d_1$ and $d_2$ are defect central charges, where $d_1$ is related to the coefficient of the displacement operator 2-point correlator and $d_2$ is related to $h$ in \eqref{Tbulk} through \eqref{eq-hd2}. 

Another way to see the Weyl anomalies of surface defects is that there are ultraviolet divergences in the expectation value of surface operators
\begin{equation}
    \log \langle D_\Sigma \rangle \supset \int_\Sigma \mathcal{A}_\Sigma \mathrm{vol}_\Sigma \log\ell/\epsilon\ ,
\end{equation}
with the anomaly density $\mathcal{A}_\Sigma$ given by
\begin{equation}
    \mathcal{A}_\Sigma = \frac{1}{12\pi} \left[ b R^\Sigma + d_1 \Tilde{\Pi}_{ab}^{\mu} \Tilde{\Pi}^{ab}_{\mu} - d_2 W_{ab}^{ab} \right]\ . 
\end{equation}

\section{Surface defects in free fields}
In this section we compute the surface defect contribution to bulk R\'{e}nyi entropy in free fields, and verify a previous formula for entanglement entropy \eqref{JensenOBannon}~\cite{Nishioka1,Jensen1}. We will first compute the surface defect in free scalar theory and then move to the theory of free two-form fields. The results are compared with the line defect results in four dimensions obtained in~\cite{LM1}.

We consider a free conformal scalar $\phi$ and 2-form field $B$ with surface operators introduced in six dimensions. The bulk theories are
\begin{equation}
    L_\phi = \frac{1}{2}(\partial\phi)^2 + \frac{1}{10} R\phi^2\ ,
\end{equation}
\begin{equation}
    L_B = \frac{1}{12} F_{\mu_1\mu_2\mu_3} F^{\mu_1\mu_2\mu_3} \ ,
\end{equation}
and the surface operators are
\begin{equation}
    D_\phi = \exp\left(\int_\Sigma \mathrm{d}^2\sigma\phi(\sigma)\right)\ ,
\end{equation}
\begin{equation}
    D_B = \exp\left(i\int_\Sigma B\right)\ .
\end{equation}
For these free theories, $h$ can be computed by considering a planar surface~\cite{defect2016}
\begin{equation}\label{eq-h}
h_\phi = \frac{1}{20\pi^4}\ ,\quad h_B =\frac{1}{4\pi^4} \ .
\end{equation}

\begin{figure}[htbp]
\centering
\includegraphics[height=5.1cm]{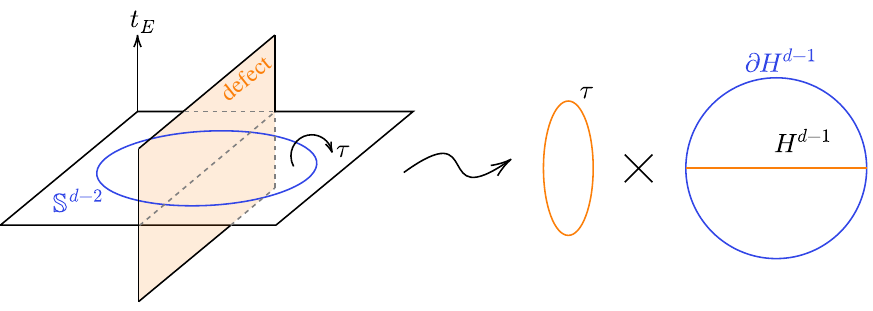}
\caption{Illustration of the map \eqref{eq-coordT}. }
\label{fig-trans}
\end{figure}

Now we are going to compute the defect contribution to bulk R\'enyi entropy by mapping the system to $S^1 \times H^5$. We start from the Euclidean flat space $\mathbb{R}^d$
\begin{equation}
    \mathrm{d} s_{\mathbb{R}^d}^2 = \mathrm{d}t_E^2 + \mathrm{d}r^2 + r^2\mathrm{d}\Omega_{d-2}^2\ ,
\end{equation}
with the surface defect along $t_E$ and $r$ and located at $\theta = 0,\pi$. The entangling surface is given by $(t_E = 0, r = \ell)$. Using the following coordinate transformation
\begin{equation}\label{eq-coordT}
    t_E = \ell\frac{\sin\tau}{\cosh\rho + \cos\tau}\ ,\quad r = \ell\frac{\sinh\rho}{\cosh\rho + \cos\tau}\ , 
\end{equation}
with $\tau\in[0,2\pi)$ and $\rho\in[0,+\infty)$, $\mathbb{R}^d$ is mapped to $S^1 \times H^5$ with a warp factor $\Omega = (\cosh\rho + \cos\tau)^{-1}$
\begin{equation}
     \mathrm{d} s_{\mathbb{R}^d}^2 = \Omega^2 \mathrm{d}s_{S^1 \times H^5}^2 = \Omega^2 \ell^2 (\mathrm{d}\tau^2 + \mathrm{d}\rho^2 + \sinh^2\rho\,\mathrm{d}\Omega_{d-2}^2)\ . 
\end{equation}
At the same time, the surface defect is mapped to $S^1\times H^1$, i.e., along $\tau$ and $\rho$ and located at $\theta=0,\pi$, as illustrated in Figure~\ref{fig-trans}. 
\subsection{R\'enyi entropy for a free scalar field}
We use heat kernel method to compute the Green function. The heat kernel of Laplacian for a free scalar on $H^5$ is given by
\begin{equation}
K_{H^5}(t,\rho) = \frac{e^{-4 t - \frac{\rho^2}{4 t}} \left(\rho^2 + 2 \rho\,t \coth\rho - 2 t\right)\text{csch}^2\rho }{32 \pi ^{5/2} t^{5/2}}\ ,
\end{equation}
where $\rho$ is the geodesic distance on $H^5$. The factor $e^{-4t}$ should be cancelled by the additional conformal mass. The heat kernel on $S^1_\beta$ is
\begin{equation}
K_{S^1_\beta}(t,\tau) = \sum_{n=-\infty}^{+\infty}\frac{\exp \left(-\frac{(\beta  n+\tau )^2}{4 t}\right)}{\sqrt{4 \pi  t}}\ ,
\end{equation}
where $\beta$ is the length of the circle and $\tau$ is the Euclidean time distance. If we set one point at $(\tau = 0, \rho = 0, \theta = 0, \theta_1 = 0, \theta_2 = 0, \varphi = 0)$ and the other point at an arbitrary place, then the propagator is given by
\begin{equation}
G_\beta(\tau,\rho) = \int_0^{+\infty} \mathrm{d}t\, K_{S^1_\beta}(t,\tau)*K_{H^5}(t,\rho)e^{4t}\ .
\end{equation}
Evaluating this integral we obtain a complicate expression for $G_\beta$,
\begin{equation}\label{eq-GreenBeta}
\frac{\mathrm{csch}^2\rho \left[\beta \coth\rho \left(\coth\frac{\pi(\rho + i\tau)}{\beta} + \coth\frac{\pi(\rho - i\tau)}{\beta}\right) + \pi\left(\mathrm{csch}^2 \left(\frac{\pi (\rho + i\tau)}{\beta}\right) + \text{csch}^2\left(\frac{\pi (\rho - i\tau)}{\beta}\right)\right)\right]}{16 \pi^2 \beta^2}.
\end{equation}
However, if we take $\beta\to 2\pi$, this becomes the familiar propagator on $S^1_{2\pi} \times H^5$
\begin{equation}\label{eq-Green2Pi}
G(\tau,\rho) = \frac{1}{16 \pi ^3 \left(\cos \tau - \cosh \rho\right)^2}\ .
\end{equation}
To calculate the R\'enyi entropy, we need the defect free energy as a function of $\beta$, which can be calculated by integrating \eqref{eq-GreenBeta} over $S^1_\beta \times H^1$. This integral diverges at $\rho = \tau = 0$. To regularize this divergence we exclude the region $\tau^2 + \rho^2 < \epsilon^2$ and end up with zero universal term (See Appendix~\ref{appx-Scalar2pt} for a detailed calculation). This means that the surface defect contribution to free energy vanishes at any temperature, similar to what happens for line defect in $4d$~\cite{LM1}. However, as pointed out in~\cite{LM1}, we should not forget the boundary contribution from the tip of the cone. 

The additional contribution to the entropy is an area-like term (the Wald term) from the conformal coupling of the scalar to curvature,
\begin{equation}\label{eq-WaldTerm}
S = -\frac{4\pi}{10}\int_{\hat{\mit\Sigma}} \mathrm{d}A \llangle\phi^2\rrangle\ ,
\end{equation} 
where $\hat{\mit\Sigma}$ is the tip of the cone located at $\rho=\infty$. Evaluating this integral, we obtain the surface defect contribution to the bulk R\'enyi entropy in free scalar theory
\begin{equation}\label{eq-WaldTermR}
S_\phi = -\frac{1}{10\pi}\log\ell/\epsilon\ ,
\end{equation}
which does not rely on $\beta$. A detailed calculation of \eqref{eq-WaldTermR} is given in Appendix~\ref{appx-WaldTerm}. Now let us check against a recently found formula of entanglement entropy with surface defect~\cite{Jensen1},
\begin{equation}\label{JensenOBannon}
S = \frac{1}{3}\left(b - \frac{d-3}{d-1}d_2\right)\log\ell/\epsilon\ ,
\end{equation} where $d_2$ is related to $h_\phi$ by
\begin{equation}\label{eq-hd2}
d_2 = 6\pi\Omega_{d-3}\frac{d-1}{d}\, h_\phi\ ,
\end{equation}
with $\Omega_{d-3}$ the area of $d-3$ dimensional sphere. From our heat kernel analysis we have $b=0$ (this is consistent with~\cite{Drukker1,Cuomo:2023qvp}). Plug in $h_\phi = 1/20\pi^4$ and $d = 6$, we find that our entropy result is perfectly consistent with \eqref{JensenOBannon}.
\subsection{R\'enyi entropy for a free 2-form field}
Having verified the relation between entanglement entropy and defect central charges $b$ and $h$, one can compute the entanglement entropy for a free 2-form field in six dimensions. The $b$ central charge comes from the logarithmic divergence term in the expectation value of spherical defect. This was essentially computed in~\cite{Drukker1}\footnote{See~\cite{Henningson:1999xi,Gustavsson:2003hn,Gustavsson:2004gj} for early calculations.} and in our normalization it is given by
\begin{equation}
b = \frac{3}{2\pi}\ ,
\end{equation}
where a factor of $2$ was included since we are dealing with a 2-form field without reducing half of its degrees of freedom. Plug in $h_B = 1/4\pi^4$ and $d = 6$, we have
\begin{equation}
S_B = \frac{1}{3}\left(b - 6\pi^3 h_B\right)\log\ell/\epsilon = 0\ .
\end{equation}
This result indicates that the expectation value of surface defect with $2$-form on $S^1_\beta \times H^5$ is proportional to $\beta$, which is similar to line defect in $4d$ Maxwell theory~\cite{LM1}. Furthermore, from information theory, R\'enyi entropy also vanishes. It is interesting to demonstrate this result by heat kernel computation, which we leave for future work.
\section{\boldmath Surface defects in large $N$ limit\unboldmath}
Let us now move to the most famous six dimensional theory, the $(2,0)$ theory, and apply the ideas in the previous sections to the well-known observable, the surface operator, which has been studied both from field theory perspective~\cite{Henningson:1999xi,Gustavsson:2003hn,Gustavsson:2004gj} and from holography~\cite{Berenstein:1998ij,Graham:1999pm}. A six-dimensional $(2,0)$ tensor multiplet includes $5$ real scalars, $2$ Weyl fermions, and a $2$-form field with self-dual strength, which can be considered as a chiral 2-form field with half of the degrees of freedom. In the free theory with a tensor multiplet, the surface operator can be defined in analogy to Maldacena-Wilson loop,
\begin{equation}
W = \exp \int_\Sigma (i B^+ - n^i \phi_i \mathrm{vol}_\Sigma)\ .
\end{equation}
As a simple example, the surface operator may include (the pullback of) a chiral $2$-form and a real scalar. Therefore the defect contribution to the bulk R\'{e}nyi entropy is given by the sum
\begin{equation}
S = S_\phi + \frac{1}{2} S_B = -\frac{1}{10\pi}\log\ell/\epsilon\ .
\end{equation}

It was conjectured that the $A_{N-1}$ $(2,0)$ SCFT is dual to M-theory on $\text{AdS}_7\times S^4$ with $N$ units of 4-form flux on $S^4$~\cite{Maldacena:1997re},
\begin{equation}
\mathrm{d}s_{11}^2 = L^2(\mathrm{d}s^2_{\text{AdS}_7} + \frac{1}{4}\mathrm{d}s^2_{S^4})\ ,\quad F_4 = \mathrm{d}C_3 = \pi^2 L^3 \mathrm{vol}_{S^4}\ ,\quad L^3 = 8\pi N \ell_p^3\ .
\end{equation}

Due to the lack of an intrinsic definition of interacting $(2,0)$ theory and having only $N$ as a free parameter, it is unclear how to explicitly define non-trivial observables, such as surface defect, to test this AdS/CFT duality. Nevertheless, it is believed that there exist half-BPS surface defects in $(2,0)$ theories, characterized by a weight vector $\lambda\in \Lambda_w(\mathfrak{g})$, where $\mathfrak{g}$ is the ADE Lie algebra labelling the theory. We will focus on $A_{N-1}$ $(2,0)$ theories in the large $N$ limit. The holographic dual of the surface operator in fundamental representation is a single M2-brane.

To introduce an extra parameter one may consider to put $(2,0)$ theory on $S_\beta^1\times H^5$. One strong motivation to consider the theory on this manifold is to compute the R\'{e}nyi entropy in flat space. The dual M-theory background may then have the AdS$_7$ part with the corresponding $S_\beta^1\times H^5$ boundary. However it was pointed out in~\cite{Nian:2015xky} that, supersymmetries are broken for $\beta\neq 2\pi$ since there are no surviving Killing spinors in this generic background. To preserve supersymmetry, one can turn on an extra R-symmetry background field, which leads to the observable of supersymmetric R\'{e}nyi entropy~\cite{Nian:2015xky,Zhou:2015kaj}.

Given the R-symmetry twist background on the boundary, the natural holographic dual is the seven dimensional gauged $SO(5)$ supergravity, which can be obtained by Kaluza-Klein reduction of $11d$ supergravity on $S^4$~\cite{Cvetic:1999xp}. It is sufficient to consider only the action for the remaining fields after truncation, which include the metric, two gauge fields and two scalars,
\begin{equation}
    \frac{1}{\sqrt{g}} \mathcal{L} = R - \frac{1}{2}\left(\partial\Vec{\phi}\right)^2 - \frac{4}{L^2} V - \frac{1}{4}\sum_{i=1}^2 \frac{1}{X_i^2}\left(F^i\right)^2\ ,
\end{equation}
where $\Vec{\phi}=(\phi_1,\phi_2)$ are two scalars and
\begin{equation}
    X_i = e^{-\frac{1}{2}\Vec{a}_i\Vec{\phi}}\ ,\quad i=1,2\ ,\quad \Vec{a}_1 = (\sqrt{2},\sqrt{2/5})\ ,\quad \Vec{a}_2 = (-\sqrt{2},\sqrt{2/5})\ .
\end{equation} 
The potential is $V = -4X_1 X_2 - 2X_0 X_1 - 2X_0 X_2 + \frac{1}{2}X_0^2$, with $X_0 = (X_1 X_2)^{-2}$.
This theory has a 2-charge $7d$ topological black hole solution, which asymptotes to hyperbolically sliced AdS$_7$,
\begin{align}
&\mathrm{d}s_7^2 = -\left(H_1 H_2\right)^{-4/5} f \mathrm{d}t^2 + \left(H_1 H_2\right)^{1/5} \left(f^{-1} \mathrm{d}r^2 + r^2 \mathrm{d}\Omega_{5,k}^2\right)\ ,\\
&f(r) = k - \frac{m}{r^4} + \frac{r^2}{L^2} H_1 H_2\ ,\quad H_i = 1 + \frac{q_i}{r^4}\ ,
\end{align}
together with scalars and gauge fields,
\begin{equation}
X_i = \left(H_1 H_2\right)^{2/5}H_i^{-1}\ ,\quad A^i = \left(\sqrt{k} \left( H_i^{-1} - 1 \right) + \mu_i \right)\mathrm{d}t\ .
\end{equation} 
For our purpose we consider $k=-1$ and $m=0$. Let us define a rescaled charge $\kappa_i = q_i/r_H^4$, then the black hole horizon can be expressed in terms of $\kappa_i$,
\begin{equation}
r_H = \frac{L}{\sqrt{(1 + \kappa_1)(1 + \kappa_2)}}\ .
\end{equation}
The Hawking temperature of this black hole is
\begin{equation}
T = \frac{f'(r)}{4\pi \sqrt{H_1 H_2}}\big|_{r = r_H} = \frac{1 - \kappa_1 - \kappa_2 - 3\kappa_1\kappa_2}{2\pi L(1 + \kappa_1)(1 + \kappa_2)}\ .
\end{equation}
By matching to the boundary temperature $1/\beta$, one can solve $\kappa$ and therefore solve the black hole.
The Bekenstein-Hawking entropy and the holographic supersymmetric R\'{e}nyi entropy were computed in~\cite{Zhou:2015kaj}. In this note we want to solve a probe M2-brane in this background.
\subsection{Holographic surface defect} 
We want to compute the expectation value of a surface defect wrapping on $(\tau,\rho)$ direction using the relationship
\begin{equation}
S_{\text{M2}} = -\log \langle W\rangle_n \ ,
\end{equation}
where the M2-brane action is
\begin{equation}
S_{\text{M2}} = T_2 \int \mathrm{d}^3\sigma \sqrt{-\mathrm{det}[g]}\ .
\end{equation}
In the probe limit, the M2-brane solution is given by
\begin{equation}
\sigma_0 = \tau\ ,\ \sigma_1 = \rho\ ,\ \sigma_2 = r\ ,
\end{equation}
and the on-shell action is 
\begin{equation}
S_{\text{M2}} = T_2\int_0^\beta \mathrm{d}\tau \int_{-\infty}^\infty \mathrm{d}\rho \int_{r_H}^\Lambda r_H(H_1H_2)^{-1/5}\sqrt{\Delta}\ ,
\end{equation}
where $\Delta^{1/3}$ is the warp factor in front of seven-dimensional solution in the eleven-dimensional uplift. In~\cite{Cvetic:1999xp},  $\Delta$ is specified to be 
\begin{equation}
\Delta = X_0\mu_0^2 + X_1\mu_1^2 + X_2\mu_2^2\ ,
\end{equation} where $\mu_{0,1,2}$ are related to 2-sphere angles by
\begin{equation}
\mu_0 = \sin\theta\ ,\ \mu_1 = \cos\theta\sin\phi\ ,\ \mu_2 = \cos\theta\cos\phi\ .
\end{equation}
As explained for string embedding in~\cite{Crossley:2014oea}, the M2-brane should sit at the point on the internal manifold in order to preserve the R-symmetry (twist part).
Under this condition we find
\begin{equation}
S_{\text{M2}} = -2\pi n T_2 V_{H^1} r_H^2\ .
\end{equation}
For a single charged black hole with only $\kappa_1\neq 0$, which corresponds to the field theory twisting by a single $U(1)$ Cartan of the R-symmetry, the M2-brane embedding can be chosen at $\mu_1 =0$, and we find
\begin{equation}
r_H^2 = \frac{n+1}{2n}\ ,\quad S_{\text{M2}} = -2\pi T_2 V_{H^1}\left(\frac{n+1}{2}\right)\ .
\end{equation} 
A consistent check: When $n=1$, we have
\begin{equation}
S_{\text{M2}} = -2\pi T_2 V_{H^1} = -4N\log\ell/\epsilon\ ,
\end{equation} which agrees with the result in~\cite{Berenstein:1998ij,Drukker1}. The surface defect contribution to supersymmetric R\'enyi entropy is
\begin{equation}
S_n = \frac{\log\langle W\rangle_n - n \log\langle W\rangle_1}{1-n} = \pi T_2 V_{H^1}\ ,
\end{equation} which is independent of $n$.

For two equally-charged black hole, which corresponds to twisting by two $U(1)$ Cartans of the R-symmetry, the M2-brane embedding can be chosen at $\theta=\pi/2$, and we find
\begin{equation}
r_H^2 = \frac{(3n + 1)^2}{16n^2}\ ,\quad S_{\text{M2}} = -2\pi T_2 V_{H^1}\left(\frac{(3n + 1)^2}{16n}\right)\ .
\end{equation} The surface defect contribution to supersymmetric R\'enyi entropy is
\begin{equation}
S_n = 2\pi T_2 V_{H^1}\left(\frac{7n+1}{16n}\right)\ .
\end{equation}

\section{Discussion}
In this note we developed a method to compute the surface defect contribution to bulk R\'{e}nyi entropy. We mainly focus on surface defect in six dimensions but the method is applicable in other dimensions as well. We obtained explicit results for free fields and for $(2,0)$ theories in the large $N$ limit. For free fields we employ the heat kernel method in $S^1_\beta\times H^5$ with defect wrapped on $S^1_\beta\times H^1$. For large $N$ $(2,0)$ theories, we use M2-brane action in the verified supergravity solution to compute the surface defect contribution to the supersymmetric R\'{e}nyi entropy.

One interesting question is if we can make a conjecture for all $N$.
Recall that in the absence of the defect, this was achieved for supersymmetric R\'{e}nyi entropy in all known $(2,0)$ theories. The closed formula of supersymmetric R\'{e}nyi entropy connects conformal anomalies, 't Hooft anomalies, supersymmetric Casimir energy and also holography in a compact way~\cite{Zhou:2015kaj}. In the same spirit, one may hope that there exists a closed formula for the surface defect contribution to the supersymmetric R\'{e}nyi entropy.

Another interesting question is about the M2-brane in the hyperbolic black hole. By now we only count the classical M2-brane contribution to the entropy. It is interesting to go further to compute the quantum fluctuations in the M2-brane worldvolume. As corrections to the entropy this is expected to be related to the all $N$ result in a certain way.




\acknowledgments
We thank Nadav Drukker for useful discussion and for reading the manuscript. This work is supported by NSFC grant 12375063. YZ is also supported by NSFC 12247103 through Peng Huanwu Center for Fundamental Theory.

\appendix
\section{\boldmath Integral of \eqref{eq-GreenBeta} over $S^1_\beta\times H^1$\unboldmath}\label{appx-Scalar2pt}
Due to the symmetry between $\theta = 0$ and $\theta = \pi$ we only need to evaluate the integral over $\tau \in [0,\beta)$, $\rho \in [0,+\infty)$ and $\theta = 0$, with the region $\tau^2 + \rho^2 < \epsilon^2$ excluded. We first consider the region $\rho > \epsilon$. One fact about the Green function $G_\beta(\tau,\rho)$ is 
\begin{equation}
\quad \int_0^\beta \mathrm{d}\tau\, G_\beta(\tau,\rho > 0) = \frac{\coth\rho\,\csch^2\rho}{8\pi^2}\ , 
\end{equation}
thus we have
\begin{equation}\label{eq-region1}
\int_\epsilon^{+\infty} \mathrm{d}\rho \int_0^\beta \mathrm{d}\tau\, G_\beta(\tau,\rho) = \frac{1}{16\pi^2 \epsilon^2} - \frac{1}{48\pi^2} + \mathcal{O}\left(\epsilon^2\right)\ . 
\end{equation}
Now we consider the region $\rho \leqslant \epsilon$. We introduce an angle parameter $\xi$ and parameterize $\rho$ to $\epsilon\sin\xi$, thus we have
\begin{equation}\label{eq-region2}
\int_{\xi = 0}^{\pi/2} \mathrm{d}(\epsilon\sin\xi) \int_{\epsilon\cos\xi}^{\beta - \epsilon\cos\xi}\mathrm{d}\tau\, G_\beta(\tau,\epsilon\sin\xi) = \frac{1}{16\pi^2 \epsilon^2} + \frac{1}{48\pi^2} + \mathcal{O}\left(\epsilon^2\right)\ . 
\end{equation}
Adding \eqref{eq-region1} and \eqref{eq-region2}, we can see the exact cancellation of the universal terms, giving zero $b$ central charge and entanglement entropy. 

\section{The Wald term}\label{appx-WaldTerm}
In this appendix we give a detailed calculation of \eqref{eq-WaldTermR}. Since we are considering the free fields, $\llangle \phi^2 \rrangle = \llangle \phi \rrangle^2$. We first evaluate $\llangle \phi(x) \rrangle$, with $x = (0, \rho = y_c, \theta, \theta_1, \theta_2, \varphi)$ located at $\partial H^5$ and a point on defect is parameterized to $x' = (\tau, y, 0, 0, 0, 0)$, thus we have\footnote{For convenience, here we take $y\in(-y_c, y_c)$ and $\theta = 0$ for defect instead of $y\in[0, y_c)$ and $\theta = 0,\pi$. }
\begin{equation}\label{eq-1pt}
    \llangle \phi(x) \rrangle = \int_{S^1_\beta \times H^1} \mathrm{d}x' \langle \phi(x) \phi(x') \rangle = \int_{-y_c}^{y_c} \mathrm{d}y \int_0^{\beta} \mathrm{d}\tau G_\beta(\tau,r), 
\end{equation}
where $r$ is the geodesic distance between $x$ and $x'$, 
\begin{equation}
    \cosh r = \cosh y_c \cosh y - \sinh y_c \sinh y \cos \theta\ . 
\end{equation}
Evaluating the integral in \eqref{eq-1pt} gives
\begin{equation}
\llangle \phi(x) \rrangle = \frac{\coth y_c \csc^2 \theta\,\csch^2 y_c}{8\pi^2}  \left( \frac{\cos (\theta/2)}{\sqrt{\cos^2 (\theta/2) + \csch^2 y_c}}  + \frac{\sin(\theta/2)}{\sqrt{\sin^2 (\theta/2) + \csch^2 y_c}}\right) \ . 
\end{equation}
Next we integrate $\llangle \phi(x) \rrangle^2$ over $\partial H^{d-1}$, i.e.
\begin{equation}
    \int_0^{2\pi} \mathrm{d}\varphi \int_0^{\pi} \mathrm{d}\theta_2 \int_0^{\pi} \mathrm{d}\theta_1 \int_0^{\pi} \mathrm{d}\theta \sqrt{\mathrm{det}[g]} \llangle \phi(x) \rrangle^2\ , 
\end{equation}
where $\sqrt{\mathrm{det}[g]} = \sinh^4 y_c \sin^3 \theta \sin^2 \theta_1 \sin \theta_2$. Taking the limit $y_c \rightarrow +\infty$ and introducing $\epsilon/\ell$ cut-offs at $\theta = 0,\pi$ to regularize the divergence, we end up with
\begin{equation}\label{eq-WTR}
    \int_\Sigma \mathrm{d}A \llangle\phi^2\rrangle\ = \Omega_3 \frac{1}{8\pi^4} \log\frac{\ell}{\epsilon} = \frac{1}{4\pi^2} \log\frac{\ell}{\epsilon}\ . 
\end{equation}
Combining \eqref{eq-WTR} and \eqref{eq-WaldTerm} gives \eqref{eq-WaldTermR}.




\bibliographystyle{JHEP}
\bibliography{biblio}
\end{document}